\documentstyle[11pt,epsfig]{article}
\begin{document}
\vskip 1 true cm

\begin{center}
{\bf INFLUENCE OF PION-DELTA-HOLE CONFIGURATION ON THE}
\vskip 0.4 true cm
{\bf RHO MESON MASS SPECTRUM IN DENSE HADRONIC MATTER}
\vskip 1 true cm
 
G. CHANFRAY 
\vskip 0.3 true cm
{\it IPN-Lyon, 43 Bd. du 11 Novembre 1918,
F-69622 Villeurbanne C\'{e}dex, France.}
\vskip 0.6 true cm
P. SCHUCK
\vskip 0.3 true cm
{\it ISN-Grenoble, 53  Avenue des Martyrs, 38026 Grenoble-C\'edex, France}  
\end{center}

\vskip 2 true cm
\begin{center}
{\bf Abstract}
\end{center}
{\it We discuss medium modification of the rho meson 
originating from its coupling to in medium quasi-pions and delta-hole 
states. This medium effect leads to a marked structure of the rho meson strength
function in the $500\, MeV$ invariant mass region.  We  
suggest that this effect may provide an explanation of the structure observed 
by the DLS collaboration in the $Ca+Ca/C+C$ ratio at $1\, GeV/u$.}

\vskip 1.5 true cm
The modification of the rho meson mass spectrum at finite density 
and/or temperature is a very vividly debated subject. The interest 
for vector mesons is certainly    related to the expected relationship between 
the evolution of their properties especially their masses with the evolution
of the glue and scalar condensates and chiral symmetry restoration 
in a hot and dense environment. In addition, dilepton production in 
relativistic heavy ion collisions provides, through a vector dominance 
picture, a unique tool to study the rho meson in highly excited matter. 
Indeed, data obtained at CERN/SPS by the CERES collaboration in 
$S+Au$ at $200\, GeV/u$ and $Pb+Au$ at $158\, GeV/u$ \cite{CERES1, CERES2}
and the HELIOS collaboration \cite{HELIOS}  have revealed 
a significant amount of strength below the free rho meson peak in 
the invariant mass region of $500\, MeV$ \cite{CERES1, CERES2}.  
This led some authors 
to the conclusion that one is seing a downward shift of the rho 
meson mass induced by a QCD vacuum modification possibly intimately 
related to partial chiral symmetry restoration \cite{LI}. 
However it has been shown that 
more conventional many body mechanisms previously proposed 
in ref. \cite{BASIS, ASA, HERR} may explain the major part of the effect 
\cite{RAPP1}.  The main mechanism 
at work  is the coupling of the rho meson to states made of collective 
pion branch and transverse delta-hole states (ie those $\Delta-h$ states 
excited in photo-reactions) \cite{BASIS, RAPP1}. 
This mechanism alone cannot explain the whole 
effect but inclusion of the p-wave coupling of the rho to some 
higher nucleon or delta resonances, building up the so-called rhosobar
\cite{PIR}, 
can give a reasonable agreement with data when incorporated in 
realistic calculation \cite{RAPP2}. More recently, new DLS \cite{PORTER} 
data of dielectron cross section 
in nucleus-nucleus collisions at $1\,GeV$ have shown some striking features.
In particular, the $Ca+Ca$ to $C+C$ ratio exhibits a marked structure in the
$500-600 MeV$ region. The purpose of this paper is to show that 
the original approach developped in ref. \cite{BASIS} for comparison 
with previous DLS data \cite {DLS0} (and subsenquently adapted to 
the CERN regime \cite{RAPP1}) may provide, at least qualitatively, 
an explanation of the effect.  \par

We will briefly summarize 
the formalism  for the calculation of the in medium rho 
meson propagator, incorporating coupling to collective pionic modes, 
longitudinal and transverse $\Delta-h$ states and also $2p-2h$ configurations.
Then  we will present some results for the ratio of the imaginary 
part of the rho meson propagator taken at two different densities and 
 give a semi-quantitative discussion of the effect seen in the $Ca+Ca$ 
to $C+C$ ratio.

\vskip 1 true cm
The rho meson propagator at momentum $q(q^0,\bf{q})$
\begin{equation}
G(q)=q^2-m^2_\rho-\Sigma(q)
\end{equation}
is fully known once the self-energy $\Sigma(q)$ is known. In the model 
described in \cite{BASIS} this self energy is obtained by substracting 
the standard two-pion loop at $q=0$ ensuring the proper normalization 
$F_\pi(0)=1$ of the pion electromagnetic form factor. In free space it reads~:
\begin{equation}
\Sigma(q)={4 g^2\over 3}\,\int {d^3t\over (2\pi)^3}\, v^2(t)\,
{q^2\over 4\omega_t^3}\,{1\over q^2-4\omega^2_t + i\eta}
\end{equation}
with $\omega_t=(t^2+m^2_\pi)^{1/2}$. The coupling coupling constant $g=5$ 
and the parameter entering the form factor $v(t)$ have been fitted 
on the pion electromagnetic form factor and the $I=J=1$ $\pi\pi$ 
phase shifts.

In the medium the rho self energy is modified by the dressing of the 
pion propagator through p-wave coupling to $\Delta-h$ states (fig. 1a)
and various vertex correction (fig. 1b, c) necessary to ensure gauge invariance. 
In a two-level model the final result is entirely expressible in terms 
of the energy $\Omega_{1,2}(k)$ of two longitudinal collective modes built on 
pion and $\Delta-h$ states and two transverse modes with energy
${\cal E}_{1,2}(k)$. Here, we limit ourselves to the case where ${\bf q}=0$~:
\begin{equation} 
\Sigma(q_0) =\Sigma_{LL}(q_0)\,+\,\Sigma_{LT}(q_0)
\end{equation}
The first piece only depends on longitudinal modes~:
$$
\Sigma_{LL}(q_0)={4 g^2\over 3}\,\int\,{d^3k\over (2\pi)^3}\,k^2 v^2(k)
\,\sum_{i=j=1}^2\,{q_0^2\over 2\big(\Omega_i(k)+\Omega_j(k)\big)\,
\Omega_i(k)\Omega_j(k)}
$$
\begin{equation}
\times\,\bigg(1+{1\over 2}\left(\alpha_i(k) +\alpha_j(k)\right)\bigg)^2\,
{Z_i(k) \,Z_j(k)\over q^2_0- \big(\Omega_i(k) + \Omega_j(k)\big)^2\,+\,i\eta}
\label{SLL}
\end{equation}
The eigenmodes $\Omega_1(k)$ are solution of the dispersion equation~:
\begin{equation} 
\Omega_i^2(k)-\omega_k^2-k^2 \tilde\Pi^0\big(\Omega_i(k),{\bf k}\big)=0
\end{equation}
where $\tilde \Pi^0 (k^0, {\bf k})$ 
is the pion self-energy dominated by the p-wave 
coupling to the $\Delta-h$ states screened by short range correlations 
associated to the $g'=0.5$ parameter. The explicit expression of $\Omega_i(k)$ 
and the associated weigth factors $Z_i(k)$ in the two-level model can be found
in ref. \cite{AOUSS}. From eq. (\ref{SLL}), it can be seen that the 
$\rho-\pi\pi$ vertex is modified by the presence of the $\alpha_i(k)
=\tilde \Pi^0(\Omega_i(k),{\bf k})$ factors. This vertex correction 
is represented by the diagram of fig. 1b and its net effect is to kill 
 the structure at $q_0=2m_\pi$ originating from the dressing 
 of the pions in the medium (fig. 1a) ; a detailed discussion can be found 
 in \cite{BASIS}.
 
\noindent
The second piece of the in-medium self energy involves the coupling of the rho 
to the above longitudinal pionic modes and to a transverse mode, with energy
${\cal E}_i$ and strength $Y_i$, made of transverse delta-hole states and rho .
In practice the collectivity of this states is very weak and the main 
contribution comes from the first state  which is almost a pure 
transverse $\Delta-h$ state ($Y_1\simeq 1$). $\Sigma_{LT}$ reads~:
$$
\Sigma_{LT}(q_0)={4 g^2\over 3}\,\int\,{d^3k\over (2\pi)^3}\,C(k)\, v^2(k)
\,\sum_{i=j=1}^2\,{q_0^2\over 2\big(\Omega_i(k)+{\cal E}_j(k)\big)\,
\Omega_i(k){\cal E}_j(k)}
$$
\begin{equation}
\times\,
{Z_i(k) \bigg(1-Y_j(k)\bigg)
\over q^2_0- \big(\Omega_i(k) + {\cal E}_j(k)\big)^2\,+\,i\eta}
\label{SLT}
\end{equation}
with $C(k)$ given by 
$C(k)=(8/9)(f^*_{\pi N\Delta}/m_\pi)^2{\cal E}_{\Delta k}
\Gamma^2_{\pi N\Delta(k)}\rho$ with notation of ref. \cite{BASIS}. The opening 
of this channel, depicted in fig. 1c, leads to a rather marked 
structure in the invariant 
mass region $q_0=M\ge m_\pi+\omega_\Delta\simeq 500\,MeV$ (see fig.2 and discussion below). 
This is the mechanism proposed in ref \cite{RAPP1} to explain an important part
of the observed enhancement in the CERES data. We will discuss in the following 
to which extent it can provide an explanation of the structure seen in 
the $Ca+Ca/C+C$ ratio measured by the DLS collaboration.\par

In practice the above results for the self-energy has to be improved 
by the inclusion of the width of  the states. 
For instance, the width of the quasi-pion states 
is incorporated through the replacement~:
\begin{equation}
\Omega_i(k)\to \Omega_i(k)\,+i\,
{{\bf k}^2 \,Im \tilde\Pi^0(\Omega_i(k), {\bf k})\over  2\,\Omega_i(k)} 
\label{REPL}
\end{equation}
and similar replacements for the transverse states. Here, the imaginary part 
of the irreducible pion self-energy $Im \tilde\Pi^0$, calculated along the 
dispersion line $i$, takes into account the delta width corrected 
from Pauli blocking \cite{OSET} together with extra $2p-2h$ contributions 
not reducible to a delta width \cite{BASIS}. The imaginary part of the 
self-energy $\Sigma_{LL}$ and $\Sigma_{LT}$
involving two-state propagators are now calculated 
using a spectral representation from which the real part is obtained 
by dispersion relation. All the details and phenomenological input
are given in ref. \cite{BASIS}.\par

\vskip 0.5 true cm
The dilepton production is directly propotionnal to the strength function
\begin{equation}
S(q)=-{Im \Sigma(q)\over |q^2-m^2_\rho-\Sigma^(q)|^2}
\label{DILEP}
\end{equation}
This imaginary part of the rho meson propagator is shown on fig. 2 
for several densities. The bump between  $500$ and $600\, MeV$  
reflects the the pion-transverse
delta-hole intermediate states (see fig 1.c). In heavy ion reactions around 
$1 GeV/u$  incident energies, one can expect compressions of nuclei 
ranging from 
$1$ to $2.5  \rho_0$ depending on the size of the system. To see 
how the structure at $ 500- 600\, MeV$ evolves as a function of density, 
we show 
in figure 3 the ratio of the strength function (\ref{DILEP}) taken 
at two different densities typical of the $Ca+Ca$ and $C+C$ reactions 
measured by the DLS collaboration (this ratio is actually multiplied 
by the ratio of $A_p.A_t$ values corresponding to $Ca+Ca/C+C$). 
We also show on 
the same figure the DLS data. Of course this comparison can only give a 
first indication of the relevance of the mechanism mentionned just 
above. A realistic comparison requires a dynamical calculation incorporating 
precise experimental acceptance but we believe that the mechanism associated 
to $\pi-(\Delta-h)_T$ intermediate states, 
 is certainly one important  ingredient 
to account for the observed structure.
\vskip 1 true cm
Aknowledgement~: Discussion with R.J. Porter, R. Rapp and J. Wambach are 
greatfully aknowledged.    
\vfill\eject

\vfill\eject

\large{\bf FIGURE CAPTIONS}
\vskip 1 true cm
\noindent
{\bf Figure 1}~: Medium corrections to the rho meson propagator in the medium. 
(1a)~: dressing of the pions in the medium. (1b)~: vertex correction required 
by gauge invariance. (1c)~: coupling of the rho to quasi-pion and transverse
delta-hole states.
\vskip 1 true cm
\noindent
{\bf Figure 2}~: Imaginary part of the rho meson propagator at  
values of $\rho / \rho_0=0,1,2$ with imaginary part from $\Delta$ 
resonance and $2p-2h$  included.
\vskip 1 true cm
\noindent
{\bf Figure 3}~: Ratio of the strength function (multiplied by the 
ratio of $A_p.A_t$ values for $Ca+Ca/C+C$) for various couples of $\rho/\rho_0$.
Dotted line~: $1.8/ 1.2$; dashed line~: $2/1$; full line~: $2.4/1.2$.
Experimental data obtained by the DLS collaboration \cite{PORTER} 
are also shown 

\vfill\eject

\begin{figure}
  \begin{center}
    \mbox{\epsfxsize=9cm
          \epsfbox{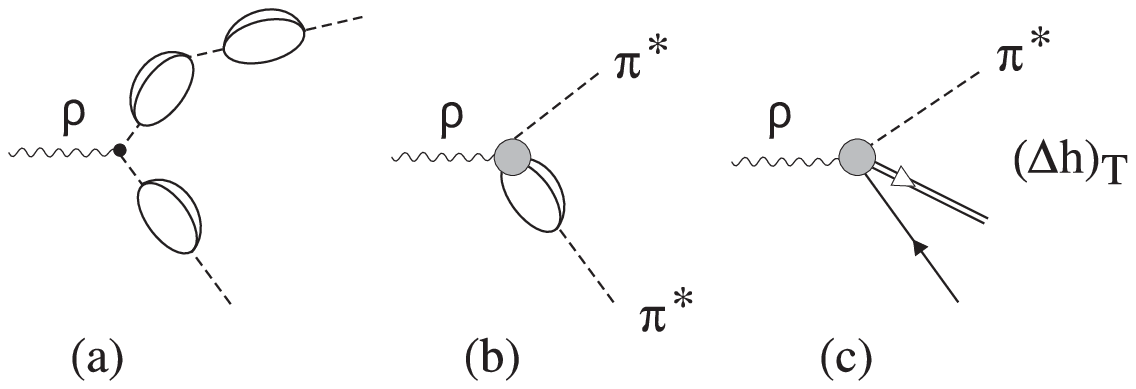}}
  \end{center}
  \caption{ }
    \label{figure1}
\end{figure}

\begin{figure}
  \begin{center}
    \mbox{\epsfxsize=9cm
          \epsfbox{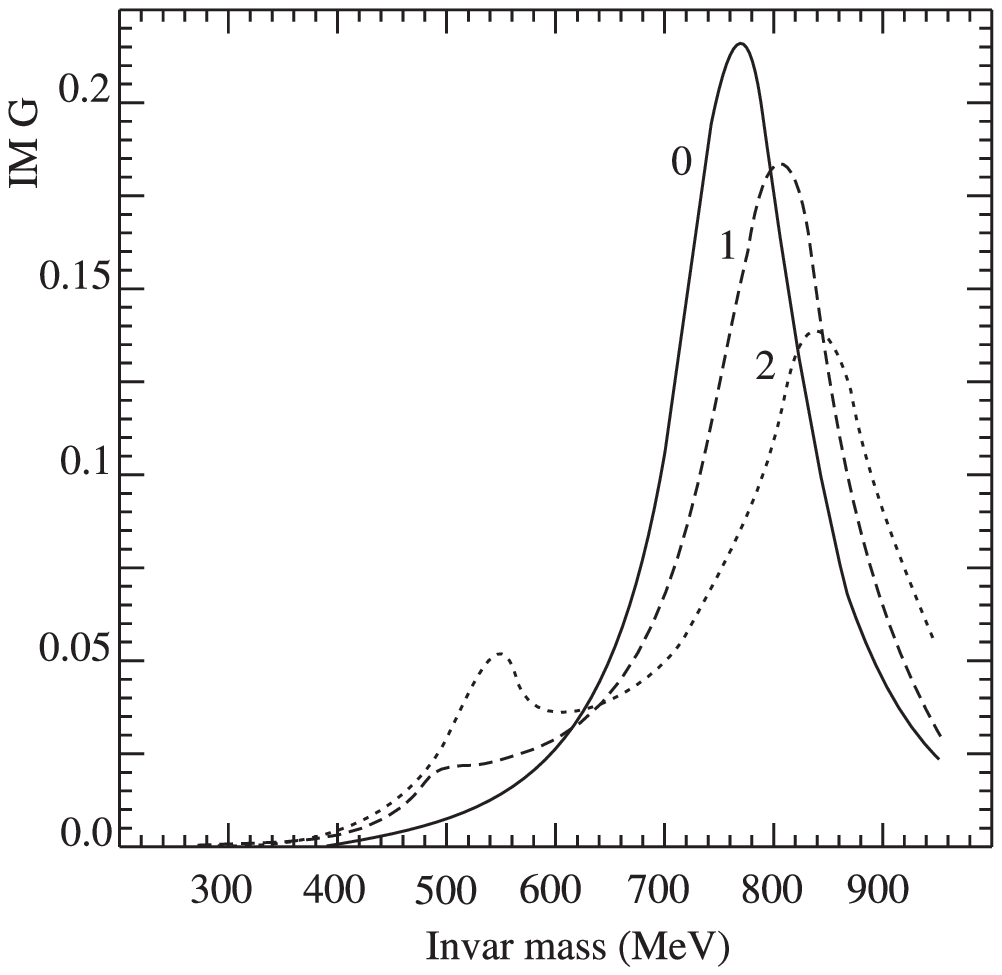}}
  \end{center}
  \caption{ }
    \label{figure2}
\end{figure}

\vfill\eject
\begin{figure}
  \begin{center}
    \mbox{\epsfxsize=12cm
          \epsfbox{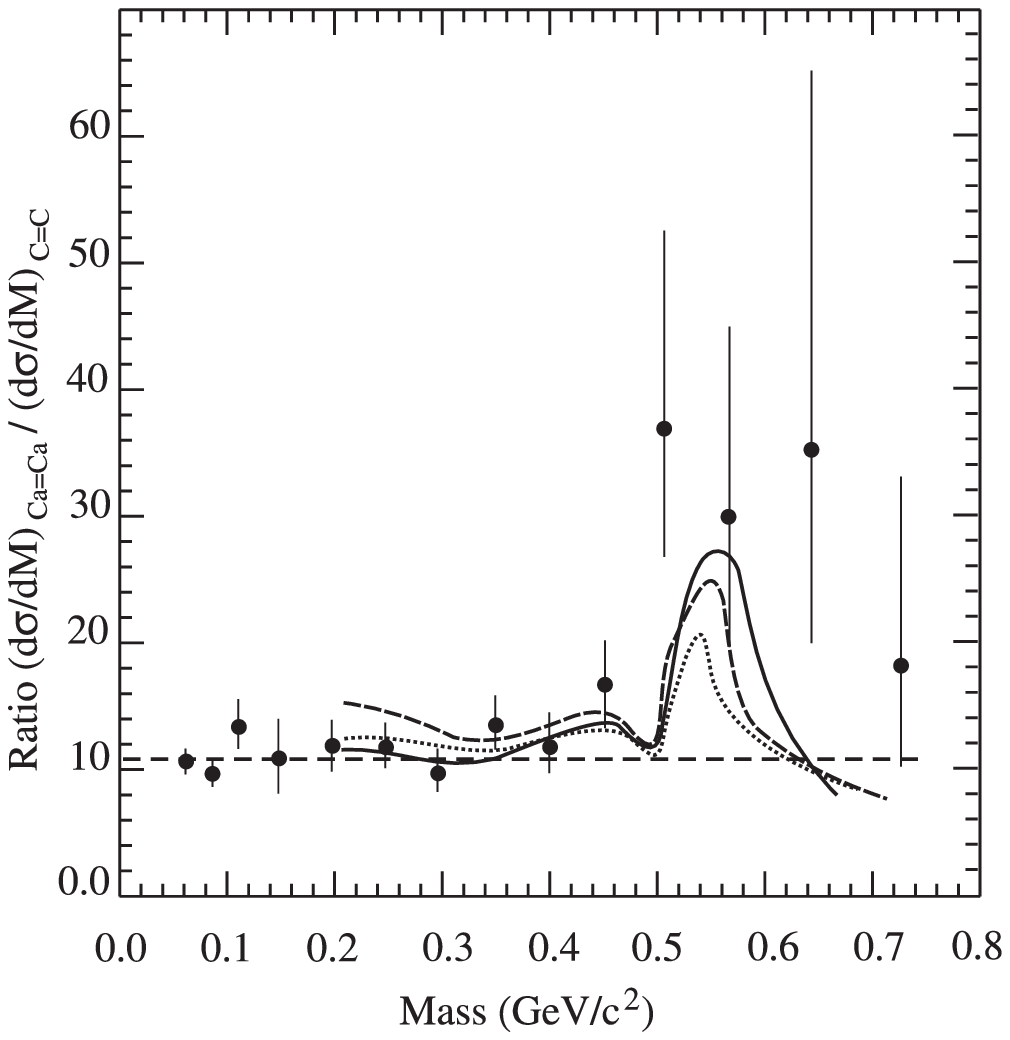}}
  \end{center}
  \caption{ }
    \label{figure3}
\end{figure}
 

\begin{thebibliography}{99}
\bibitem{CERES1} G. Agakichiev {\it et.al},  Phys. Rev. Lett. {\bf 75}, 
1272 (1995); CERES Collaboration, J.P. Wurm, Nucl. Phys. {\bf A590}, 103c 
(1995)
\bibitem{CERES2} A. Drees, Proceedings of the International Workshop on Gross 
Properties of Nuclei and Nuclear Excitations, Hirschegg, Austria January 1997, 
Editors H. Feldmeier, W. N\"{o}renberg
\bibitem{HELIOS} M. Masera {\it et.al.}, Nucl.Phys. {\bf A 590}, 93c (1995) 
\bibitem{LI} G.Q. Li, C.M. Ko, G.E. Brown, Phys. Rev. Lett. {\bf 75}, 4007
(1995)
\bibitem{BASIS} G. Chanfray, P. Schuck, Nucl. Phys. {\bf A545}, 271c (1992);
{\bf A555}, 329 (1993)
\bibitem{ASA} M. Asakawa, C.M. Ko, P. Levai, X.J. Qiu, Phys. Rev. {\bf C46},
R1159 (1992)
\bibitem{HERR} M. Herrmann, B.L. Friman, W. N\"{o}renberg,  Nucl. Phys. {\bf A545}, 
267c (1992)
\bibitem{RAPP1} G. Chanfray, R. Rapp, J. Wambach, Phys. Rev. Lett. {\bf 76}, 
368 (1996)
\bibitem{RAPP2} R. Rapp, G. Chanfray, J. Wambach, Nucl. Phys. {\bf A617},
472 (1997)
\bibitem{PIR} B. Friman, H. J. Pirner, Nucl. Phys. {\bf A617}, 496 (1997)
\bibitem{PORTER} The DLS Collaboration, R. J. Porter {\it et. al.}, 
nucl-ex/9703001 
\bibitem{DLS0} G. Roche {\it et.al}, Phys. Rev. Lett. {\bf 61}, 1069, (1988);
 Phys. Lett. {\bf B226}, 228 (1989)
\bibitem{AOUSS} G. Chanfray, Z. Aouissat, P. Schuck, W. Norenberg, 
Phys. Lett. {\bf B256}, 325 (1991)
\bibitem{OSET} E. Oset, L.L , Nucl. Phys. {\bf A468}, 631 (1987)
\end{thebibliography}
\end{document}